\documentclass[draftcls,onecolumn]{IEEEtran}
\usepackage{amsmath}
\usepackage{amssymb}
\usepackage{amsthm}
\usepackage{graphicx}
\usepackage{enumerate}
\usepackage{bbm}
\usepackage{array}
\usepackage{color}
\usepackage[table]{xcolor}
\usepackage{float}

\usepackage{hyperref}
\usepackage{cite}
\usepackage{xcolor}

\usepackage[draft]{changes}

\theoremstyle{plain}
\newtheorem{thm}{Theorem}
\newtheorem{cor}{Corollary}
\newtheorem{lem}[cor]{Lemma}
\theoremstyle{definition}
\newtheorem{exm}[cor]{Example}

\usepackage{mleftright}\mleftright
\newcommand{\dd}[1]{\xrightarrow{dd_{#1}}} 
\newcommand{\capc}{\operatorname{cap}}

\newcommand\xqed[1]{%
	\leavevmode\unskip\penalty9999 \hbox{}\nobreak\hfill
	\quad\hbox{#1}}
\newcommand\myendexm{\xqed{$\square$}}

\begin{document}
\title{Capacity and Expressiveness\\ of Genomic Tandem Duplication}
\author{\IEEEauthorblockN{Siddharth Jain}\\
\IEEEauthorblockA{Electrical Engineering\\
California Institute of Technology\\
Pasadena, CA 91125, U.S.A.\\
sidjain@caltech.edu}\\
\and
\IEEEauthorblockN{Farzad Farnoud (Hassanzadeh)}\\
\IEEEauthorblockA{Electrical Engineering\\
California Institute of Technology\\
Pasadena, CA 91125, U.S.A.\\
farnoud@caltech.edu}\\
\and
\IEEEauthorblockN{Jehoshua Bruck}\\
\IEEEauthorblockA{Electrical Engineering\\
California Institute of Technology\\
Pasadena, CA 91125, U.S.A.\\
bruck@paradise.caltech.edu}
}
\date{}
\maketitle
\begin{abstract}
The majority of the human genome consists of repeated sequences. An important type of repeated sequences common in the human genome are tandem repeats, where identical copies appear next to each other. For example, in the sequence $AGTC\underline{TGTG}C$, $TGTG$ is a tandem repeat, that may be generated from $AGTCTGC$ by a tandem duplication of length $2$. In this work, we investigate the possibility of generating a large number of sequences from a \textit{seed}, i.e.\ a small initial string, by tandem duplications of bounded length. We study the capacity of such a system, a notion that quantifies the system's generating power. Our results include \textit{exact capacity} values for certain tandem duplication string systems. In addition, motivated by the role of DNA sequences in expressing proteins via RNA and the genetic code, we define the notion of the \textit{expressiveness} of a tandem duplication system as the capability of expressing arbitrary substrings. We then \textit{completely} characterize the expressiveness of tandem duplication systems for general alphabet sizes and duplication lengths. In particular, based on a celebrated result by Axel Thue from 1906, presenting a construction for ternary square-free sequences, we show that for alphabets of size 4 or larger, bounded tandem duplication systems, regardless of the seed and the bound on duplication length, are not fully expressive, i.e.\ they cannot generate all strings even as substrings of other strings. Note that the alphabet of size 4 is of particular interest as it pertains to the genomic alphabet. Building on this result, we also show that these systems do not have full capacity. In general, our results illustrate that duplication lengths play a more significant role than the seed in generating a large number of sequences for these systems.
\end{abstract}

\begin{IEEEkeywords}
Capacity, expressiveness, tandem repeats, tandem duplication, finite automaton, irreducible strings\footnote{This paper was presented in part at IEEE International Syposium on Information Theory (ISIT), 2015}.
\end{IEEEkeywords}
\section{Introduction}\label{sec:introduction}
More than $50\%$ of the human genome consists of repeated sequences~\cite{Lander}.  Two important types of common repeats are i) interspersed repeats and ii) tandem repeats. Interspersed repeats are caused by transposons. A transposon (jumping gene) is a segment of DNA  that can copy or cut and paste itself into new positions of the genome. Tandem repeats are caused by slipped-strand mispairings~\cite{Mundy}. Slipped-strand mispairings occur when one DNA strand in the duplex becomes misaligned with the other.  

Tandem Repeats are common in both prokaryote and eukaryote genomes. They are present in both coding and non-coding regions and are believed to be the cause of several genetic disorders. The effects of tandem repeats on several biological processes is understood by these disorders. They can result in generation of toxic or malfunctioning proteins, chromosome fragility, expansion diseases, silencing of genes, modulation of transcription and translation~\cite{Usdin08} and rapid morphological changes~\cite{Fondon}. 

A process that leads to tandem repeats, e.g.\ through slipped-strand mispairing, is called \textit{tandem duplication}, which allows substrings to be duplicated next to their original position. For example, from the sequence $AGTCGTCGCT$, a tandem duplication of length $2$  can give $AGTCGT\underline{CG}\underline{CG}CT$, which, if followed by a duplication of length $3$ may give $AGTCG\underline{TCG}\underline{TCG}\penalty 0 CGCT$. The prevalence of tandem repeats and the fact that much of our unique DNA likely originated as repeated sequences~\cite{Lander} motivates us to study the capacity and expressiveness of string systems with tandem duplication, as defined below.

The model of a \textit{string duplication system} consists of a \textit{seed}, i.e.\ a starting string of finite length, a set of duplication rules that allow generating new strings from existing ones, and the set of all sequences that can be obtained by applying the duplication rules to the seed a finite number of times.  The notion of \textit{capacity}, introduced in~\cite{Farzad} and defined more formally in the sequel, represents the average number of $m$-ary symbols per sequence symbol that are asymptotically required to encode a sequence in the string system, where $m$ is the \textit{alphabet} size (for DNA sequences the alphabet size is 4). The maximum value for capacity is 1. A duplication system is \textit{fully expressive} if all strings with the alphabet appear as a substring of some string in the system. As we will show, if a system is not fully expressive, then its capacity is strictly less than~1.

Before presenting the notation, definitions, and the results more formally, in the rest of this section, we present two simple examples to illustrate the notions of expressiveness and capacity for tandem duplication string systems. Furthermore, we also outline some useful tools as well as some of the results of the paper.
\begin{exm}\label{exm:1}
Consider a string system on the binary alphabet $\Sigma=\{0,1\}$ with $01$ as the seed that allows tandem duplications of length up to 2. It is easy to check that the set of strings generated by this system start with $0$ and end with $1$. In fact, it can be proved that all binary strings of length $n$ which start with $0$ and end with $1$ can be generated by this system. The proof is based on the fact that every such  string can be written as $0^{r_1}1^{r_2}\dotsm0^{r_{v-1}}1^{r_v}$, where each $r_i \geq 1$ and $v$ is even. A natural way to generate this string is to duplicate $01$ $\frac{v}{2}$ times and then duplicate the 0s and 1s as needed via duplications of length 1.
	
{\it Expressiveness:} From the preceding paragraph, every binary sequence $s$ can be generated as a substring in this system as $0s1$. For example, although $11010$ cannot be generated by this system, it can be generated as a substring of $0110101$ in the following way: $$ 01 \rightarrow 0101 \rightarrow 010101 \rightarrow 0\underline{11010}1.$$ Hence this system is fully expressive. 
	
{\it Capacity:} The number of length-$n$ strings in this system is $2^{n-2}$. Thus, encoding sequences of length $n$ in this system requires $n-2$ bits. The capacity, or equivalently the average number of bits (since the alphabet $\Sigma$ is of size 2) per symbol, is thus equal to~1. This is not surprising as the system generates almost all binary sequences.
\myendexm\end{exm} 
	
Observing these facts for an alphabet of size $2$, one can ask related questions on expressiveness and capacity for higher alphabet sizes and duplication lengths. However, counting the number of length-$n$ sequences for capacity calculation and characterizing fully expressive systems for larger alphabets are often not straightforward tasks. In this paper, we study these questions and develop methods to answer them.
	
\begin{figure}
	\centerline{\includegraphics[width=1.4in,keepaspectratio]{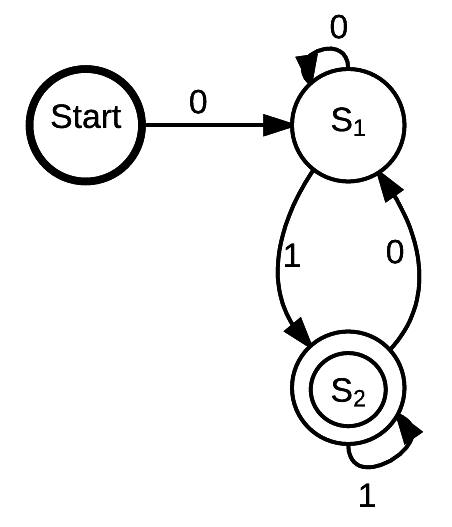}}
	\caption{Finite automaton for the systems $S = (\{0, 1\}, 01,\mathcal{T}_{\leq k}^{tan})$, where $k \geq 2$, including the system of Example \ref{exm:1}. Notation used here is described in detail in Section \ref{sec:prelim}. }
	\vspace{-1em}
	\label{fig000}
\end{figure}

A useful tool in this study is the theory of finite automata. As a simple example note that the string system over binary alphabet in the preceding example can be represented by the finite automaton given in Figure \ref{fig000}. The regular expression for the language defined by the finite automaton is
\begin{equation}\label{reg_exp_01}
	R_{01} = {(0^+1^+)}^+,
\end{equation} 
which represents all binary strings that start with $0$ and end with $1$. Here, for a sequence $s$, $s^+$ denotes one or more concatenated copies of $s$.

One can use the Perron-Frobenius theory~\cite{Immink,Lind} to count the number of sequences which can be generated by a finite automaton. This enables us to use finite automata as a tool to calculate capacity for some string duplication systems with tandem repeats over larger alphabet.

In our results, we find that the exact capacity of the tandem duplication string system over ternary alphabet with seed $012$ and duplication length at most $3$ equals $\log_3\frac{3+\sqrt{5}}{2} \simeq 0.876036$. Moreover, we generalize this result by characterizing the capacity of tandem duplication string systems over an arbitrary alphabet and a seed with maximum duplication length of 3. Namely, we show that if the maximum duplication length is 3 and the seed contains $abc$ as a substring, where $a$, $b$, and $c$ are distinct symbols, then the capacity $\simeq 0.876036\log_{|\Sigma|}3$. If such a substring does not exist in the seed, then the capacity is given by $\log_{|\Sigma|}2$, unless the seed is of the form $a^m$, in which case the capacity is $0$. Some of these results are highlighted in Table~\ref{table:cap}.

Our next example presents a system that, unlike that of Example~\ref{exm:1}, is not fully expressive.
\begin{exm} 
Consider a tandem duplication string system over the ternary alphabet $\{0,1,2\}$ with seed $012$ and maximum duplication length $3$. This system is not fully expressive as it cannot generate 210, 102, or 021, even as a substring. It is not difficult to see that to generate any of these strings, at least one of the other two must be already present as a substring of the seed. Since 012 does not contain any, by induction, it follows that the system is not fully expressive.
\myendexm\end{exm}

Based on the previous example, one may ask what happens if we start with a seed that contains one of the strings 210, 102, or 021, e.g.\ if we let the seed be 01210? Does the system become fully expressive? While this system can generate all strings of length 3 as substrings, the answer is still no as shown in Theorem \ref{thm:2}: Regardless of the seed, a ternary system with maximum duplication length of 3 is not fully expressive. We show in Theorem~\ref{thm:4}, that a maximum duplication length of at least $4$ is needed to arrive at a fully expressive ternary system.

While for alphabets of size 2 or 3, increasing the maximum length on duplications turns a system that is not fully expressive to one that is, for alphabets of size 4 or more, these systems are not fully expressive regardless how large the bound on duplication length is. The main tool in constructing quaternary strings that do not appear independently or as substrings in these systems is Thue's result proving the existence of ternary square-free sequences of any length. Note that unary and binary square-free sequences of arbitrarily large length do not exist. 
The existence of such sequences underlies the significant shift in the behavior of tandem duplication systems with regards to expressiveness as a function of alphabet size. Some of our results on expressiveness are summarized in Table~\ref{table:div}.

As part of this paper, we also study regular languages for tandem duplication string systems. In \cite{Leupold2}, it was shown that the tandem duplication string system is not regular if the maximum duplication length is 4 or more when the seed contains $3$ consecutive distinct symbols as a substring. However for maximum duplication length $3$, this question remained open. In this paper,  we show in Theorem \ref{thm:5} that if the maximum duplication length is 3, a tandem duplication string system is regular irrespective of the seed and the alphabet size.  Moreover, we characterize the exact capacity for all these systems.

\begin{table}
	\begin{center}
		\begin{tabular}{|c|c|c|c|}
			\hline
			$\Sigma$ & $s$ &$k$ & Capacity\\\hline 
			$\{0,1,2\}$ &$012$ &$3$ & $\simeq0.876036$ \\\hline
			arbitrary   & $xabcy$ &$3$&$\simeq0.876036\log_{|\Sigma|}3$\\\hline
		\end{tabular}
	\end{center}
	\caption{Capacity values tandem duplication string systems $(\Sigma, s,\mathcal{T}^{tan}_{\leq k})$. Here $x,y\in \Sigma^*$, and $a, b,c\in\Sigma$ are distinct.}
	\label{table:cap}
\end{table}

\begin{table}
	\begin{center}
		\begin{tabular}{|c|c|c|c|}
			\hline
			$\Sigma$ & $s$ &$k$ & fully expressive \\\hline
			$\{0,1,2\}$ & arbitrary & $\leq 3$ &No \\\hline
			$\{0,1,2\}$&$012$& $\geq 4$& Yes\\\hline
			Size $\geq 4$ &arbitrary &arbitrary & No \\\hline
		\end{tabular}
	\end{center}
	\caption{Expressiveness of tandem duplication string systems~$\left(\Sigma, s, \mathcal{T}^{tan}_{\leq k}\right)$.}
	\vspace{-1.5em}
	\label{table:div}
\end{table}

\paragraph*{Related Work}
Tandem duplications have already been studied in~\cite{Dassow1,Dassow2,Leupold}. However the main concern of these works is to determine the place of tandem duplication rules in the Chomsky hierarchy of formal languages. A study related to our work can be found in \cite{Farzad,Leupold2}. String systems with different duplication rules namely - end duplication, tandem duplication, reversed duplication and duplication with a gap are defined and studied in \cite{Farzad}. In end duplication, a substring of certain length $k$ is appended to the end of the previous string - for example, $A\underline{CT}GT \rightarrow ACTGT\underline{CT}$. In reversed tandem duplication, the reverse of a substring is appended in tandem in the previous string - for example, $A\underline{CT}GT \rightarrow ACT\underline{TC}GT$. In duplication with a gap, a substring is inserted after a certain gap $g$ from its position in the previous string - for example $A\underline{CT}GT \rightarrow ACTG\underline{CT}T$. 

For tandem duplication string systems, the authors in \cite{Farzad} show that for a fixed duplication length the capacity is $0$. Further, they find a lower bound on the capacity of these systems, when duplications of all lengths are allowed.  In this paper, we consider tandem duplication string systems, where we restrict the maximum size of the block being tandemly duplicated to a certain \textit{finite} length. In \cite{Leupold2}, the authors show that for these \textit{bounded} tandem duplication string systems if the maximum duplication length is $4$ or more and the alphabet size is more than 2, the system is not regular for any seed that contains 3 consecutive distinct symbols as a substring. However for maximum duplication length 3, this question was left open. In this paper, we show in Theorem \ref{thm:5} that the language is regular for maximum duplication length 3 irrespective of the seed and the alphabet size. We also characterize the exact capacity of these systems. 

 In the rest of the paper, the term tandem duplication string system refers to these kind of string duplication systems with bounded duplication length. 
 
 The rest of the paper is organized as follows. In Section \ref{sec:prelim}, we present the preliminary definitions and notation. In Section \ref{sec:capexp}, we derive our main results on capacity and expressiveness. In Section \ref{sec:reg}, we show that if the maximum duplication length is 3, then the tandem duplication string system is regular irrespective of the seed and alphabet size. Further, using the regularity of the systems, we extend our capacity results. We present our concluding remarks in Section \ref{sec:conc}.

\section{Preliminaries}\label{sec:prelim}
Let $\Sigma$ be some finite alphabet. An $n$-string $x = x_1x_2\dotsm x_n$ $\in$ $\Sigma^n$  is a finite sequence where $x_i$ $\in$ $\Sigma$ and $|x| = n$. The set of all finite strings over the alphabet $\Sigma$ is denoted by $\Sigma^*$. For two strings $x~\in~\Sigma^n$ and $y~\in~\Sigma^m$, their concatenation is denoted by $xy~\in~\Sigma^{n+m}$. For a positive integer $m$ and a string $s$, $s^m$ denotes the concatenation of $m$ copies of $s$. A string $v~\in~\Sigma^*$ is a substring of $x$ if $x = uvw$, where $u, w~\in~\Sigma^*$.

A string system $S \subseteq \Sigma^*$  is represented as a tuple $S = (\Sigma, s, \mathcal{T})$, where $s \in \Sigma^*$ is a finite length string called seed, which is used to initiate the duplication process, and $\mathcal{T}$ is a set of rules that allow generating new strings from existing ones~\cite{Farzad}. In other words, the string system $S = (\Sigma, s, \mathcal{T})$ contains all strings that can be generated from $s$ using rules from $\mathcal T$ a finite number of times. 

A tandem duplication map $T_{i,k}$,
\begin{equation*}
T_{i,k}(x)=\begin{cases}
uvvw, & \quad x=uvw,|u|=i,|v|=k,\\
x, & \quad\mbox{else},
\end{cases}
\end{equation*}
creates and inserts a copy of the substring of length $k$ which starts at position $i+1$. We use $\mathcal T^{tan}_k:\Sigma^*\to\Sigma^*$ and $\mathcal T^{tan}_{\le k}$ to denote the set of tandem duplications of length $k$, and tandem duplications of length at most $k$, respectively,
\begin{equation*}
\begin{split}
\mathcal{T}^{tan}_{k}&=\left\{ T_{i,k}:i\in\mathbb{N}\right\} ,\\
\mathcal{T}^{tan}_{\le k}&=\left\{ T_{i,j}:i,j\in\mathbb{N},j\le k\right\} .
\end{split}
\end{equation*}
With this notation, the system of Example~\ref{exm:1} can be written as  $(\{0,1\}, 01, \mathcal{T}_{\leq 2}^{tan})$.

The \textit{capacity} of the string system $S=(\Sigma,s,\mathcal T)$ is defined as
\begin{equation}
\capc(S) =\limsup_{n \rightarrow \infty} \frac{\log _{|\Sigma|}|S\cap\Sigma^n|}{n}.
\end{equation}
Furthermore, it is \textit{fully expressive} if for each $y~\in~\Sigma^*$, there exists a $z ~\in~S$, such that $y$ is a substring of $z$.


\section{Capacity and Expressiveness}\label{sec:capexp}
In this section, we present our results on the capacity and expressiveness of tandem duplication system with bounded duplication length. The section is divided into two parts; the first part focuses on capacity and the second on expressiveness.
\subsection{Capacity}
Our first result is on the capacity of a tandem duplication string system over ternary alphabet. 

\begin{thm}
	\label{thm:1}
	For the tandem duplication string system $S = \left(\{0,1,2\}, 012,\mathcal{T}^{tan}_{\leq 3}\right)$, we have \[\capc(S) = \log_3\frac{3+\sqrt5}{2}\simeq 0.876036.\]
\end{thm}

\begin{figure}
\centerline{\includegraphics[width= \columnwidth,keepaspectratio]{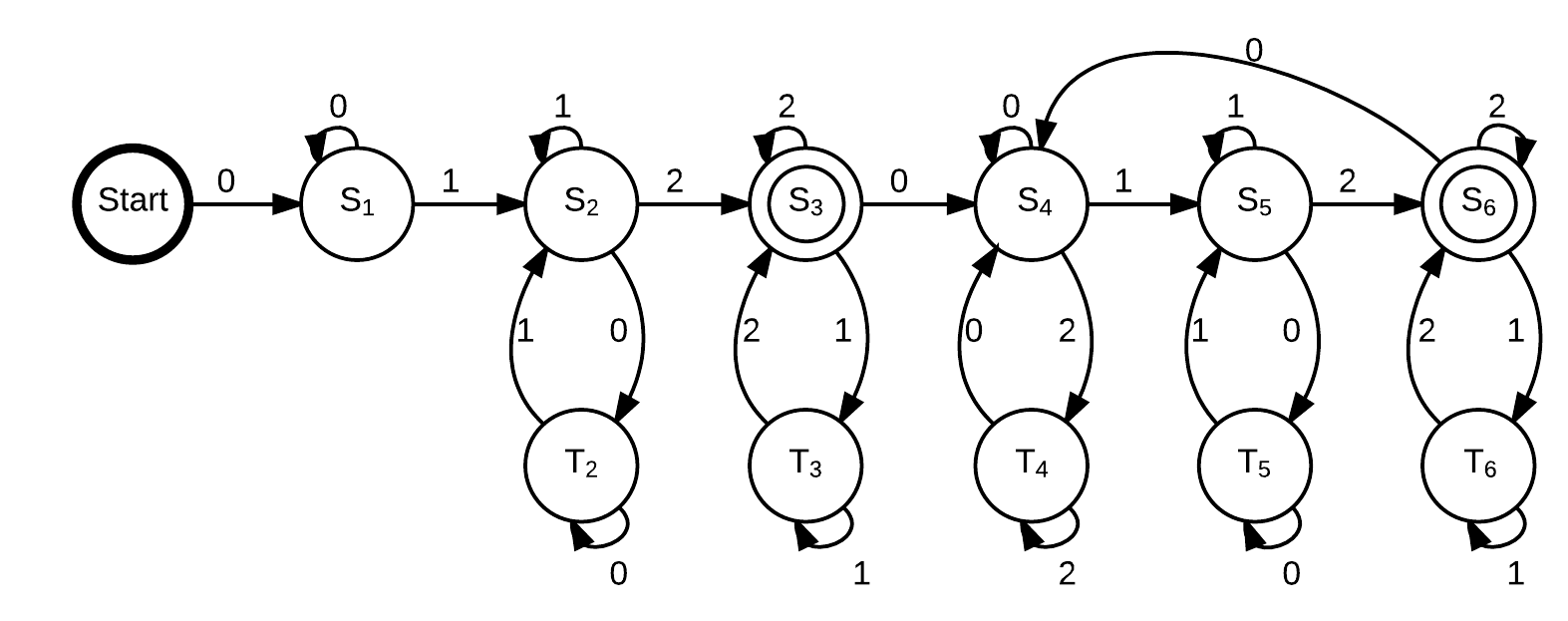}}
\caption{Finite automaton for $S = (\{0, 1, 2\}, 012,\mathcal{T}_{\leq 3}^{tan})$.}
\vspace{-2em}
\label{fig1}
\end{figure}
\begin{IEEEproof}
We prove this theorem by showing that the finite automaton given in Figure~\ref{fig1} accepts precisely the strings in $S$, and then finding the capacity using the Perron-Frobenius theory~\cite{Immink, Lind}.

The regular expression $R$ for the language defined by this finite automaton is given by
\begin{equation}\label{reg_exp}
R =  {(0^+1^+)}^+2^+{(1^+2^+)}^*{[0^+{(2^+0^+)}^*1^+{(0^+1^+)}^*2^+{(1^+2^+)}^*]}^*.
\end{equation}
Let $L_R$ be the language defined by the regular expression $R$ (and by the finite automaton). We first show that $L_R\subseteq S$. The direct way of doing so is to start with $012$ and generate all the sequences in $L_R$ via duplications. For simplicity of presentation, however, we take the reverse route: We show that every sequence in $R$ can be transformed to $012$ by a sequence \textit{deduplications}. A deduplication of length $k$ is an operation that replaces a substring $\alpha\alpha$ by $\alpha$ if $|\alpha|= k$. For two regular expressions $R_1$ and $R_2$, we use $R_1\dd{\le k} R_2$ to denote that \textit{each} sequence in $R_1$ can be transformed into \textit{some} sequence in $R_2$ via a sequence of deduplications of length at most $k$.

Note that $R = B_1{B_2}^*$, where 
\begin{equation*}\label{B1}\begin{split}
B_1 &= {(0^+1^+)}^+2^+{(1^+2^+)}^*,\\
B_2 &=  0^+{(2^+0^+)}^*1^+{(0^+1^+)}^*2^+{(1^+2^+)}^*.
\end{split}
\end{equation*} 
We have 
$B_{1}\dd{\le3}012\left(12\right)^{*}\dd{\le3}012,$ 
since $a^{+}\dd{\le3}a$ and $\left(ab\right)^{+}\dd{\le3}ab$ for all $a,b\in\Sigma$.
Furthermore,%
\begin{equation}
\label{eq:B2}
B_2\dd{\le3} 0{(20)}^*1{(01)}^*2{(12)}^* \dd{\le3} 0{(20)}^*1{(01)}^*2\dd{\le3} 0{(20)}^*12 \dd{\le3} \{02012,012\}. 
\end{equation}
Note for example that $1{(01)}^*\underline{2}{(12)}^* \dd{\le3} 1{(01)}^*2$ as the underlined $2$ is always preceded by a $1$. 

We thus have $R=B_1B_2^*\dd{\le3}\{01202012,012012\}\dd{\le3}012$, proving that $L_R\subseteq S$.


To complete the proof of $L_R=S$, we now show that $S \subseteq L_{R}$. In what follows, we say a finite automaton generates a sequence $s$, if there is a path with label $s$ from $Start$ to an accepting state. If an automaton generates $uvw$, with $u,v,w\in\Sigma^*$, we may use $v$ to refer both to the string $v$ itself and to the part of the path that generates $v$. The meaning will be clear from the context.

We show $S\subseteq L_R$, by proving the following for the finite automaton in Figure~\ref{fig1}: 

i) It can generate $012$. 

ii) If the automaton can generate $pqr$, with $p,q,r$ $\in \Sigma^*$ and $|q| \leq 3$, it can also generate $pq^2r$. 

Condition i) holds trivially (see the path $Start-S_1-S_2-S_3$ in Figure \ref{fig1}).
In order to prove ii), we define:
\begin{itemize}
\item \textit{Path Label}: Given a path $a$ in a finite automaton, the path label $l_a \in \Sigma^*$ is defined as the sequence obtained by concatenating the labels on the edges forming the path.
\item \textit{Path Length} is the number of edges of the path.
\item \textit{Superstate}: A state $D$ is a superstate of a state $C$ if for each path starting in $C$ and ending in an accepting state, there is a path with the same label starting in $D$ and ending in an accepting state. Note that every state is a superstate of itself. 
\item \textit{Duplicable Path}: A path ending in a state $C$ is \textit{duplicable} if there is a path with the same label starting in $C$ and ending in a superstate of $C$. 
\end{itemize}

Suppose a finite automaton can generate $pqr$. If $q$ is duplicable, then $pq^2r$ can also be generated by the finite automaton. As a result, to prove ii), it suffices to show that for each state $C$ in Figure \ref{fig1}, all paths of length $1$, $2$ or $3$ ending in $C$ are duplicable. 

The rest of the proof is divided into two parts. In Part 1, we show that all paths ending in $\{S_4,S_5, S_6, T_4, T_5, T_6\}$ with length $\le3$ are duplicable. In Part 2, we prove the same statement for the states $\{S_1,S_2,S_3,T_2,T_3\}$. Note that there are no nontrivial paths ending in the $Start$ state. 

{\it Part 1} : Given a state $u$ and $j\in\{1,2,3\}$, let $P^u_j$ be the set of all length-$j$ paths ending in $u$ and let $Q^u_j$ be the set of all length-$j$ paths starting and ending in $u$. If \begin{equation}\label{prop}
\bigcup_{a~\in~ P^u_j} l_a = \bigcup_{a~\in~ Q^u_j} l_a,
\end{equation}
then all length-$j$ paths ending in $u$ are duplicable.

We prove that (\ref{prop}) holds for all states $\{S_4,S_5, S_6, T_4, T_5, T_6\}$ and all $j\in\{1,2,3\}$. This is done by computing $\mathcal{A}_1$, $\mathcal{A}_1^2$ and $\mathcal{A}_1^3$, where $\mathcal{A}_1$ is the (labeled) adjacency matrix of the strongly connected component of the finite automaton given in Figure~\ref{fig1}, i.e.\ the subgraph induced by $\{S_4,S_5, S_6, T_4, T_5, T_6\}$. Here in computing the matrix products, symbols do not commute, e.g.\ $xy\neq yx$.  The adjacency matrix $\mathcal{A}_1$ and its square $\mathcal A_1^2$, where $x,~y$ and $z$ represent edges labeled by $0$, $1$, and $2$, respectively, and where rows and columns correspond in order to $S_4,S_5, S_6, T_4, T_5, T_6$, are given by 
\begin{equation*}
\mathcal{A}_1 = 
\left[\begin{smallmatrix} 
x & y & 0 & z & 0 & 0 \\
0 & y & z & 0 & x & 0 \\
x & 0 & z & 0 & 0 & y \\
x & 0 & 0 & z & 0 & 0 \\
0 & y & 0 & 0 & x & 0 \\
0 & 0 & z & 0 & 0 & y \\
\end{smallmatrix}\right],
\end{equation*}
\begin{equation*}
\vspace{1em}
\mathcal{A}_1^2 = 
\left[\begin{smallmatrix} 
x^2+zx & y^2+xy & yz & z^2+xz & yx & 0 \\
zx & y^2+xy & z^2+yz & 0 & x^2+yx & zy \\
x^2+zx & xy & z^2+yz & xz & 0 & y^2+zy \\
x^2+zx & xy & 0 & z^2+xz & 0 & 0 \\
0 & y^2+xy & yz & 0 & x^2+yx & 0 \\
zx & 0 & z^2+yz & 0 & 0 & y^2+zy 
\end{smallmatrix}\right].
\end{equation*} 
Each entry in these matrices lists the paths of specific length from the state identified by its row to the state identified by its column. For example, the entry $(6,3)$ of $\mathcal A_1^2$, which equals $z^2+yz$, indicates that there are two paths of length $2$ from $T_6$ to $S_6$ with labels $z^2=22$ and $yz=12$.

For a state $u\in\{S_4,S_5,S_6,T_4,T_5,T_6\}$, the terms in the column that corresponds to $u$ in these matrices represent the labels of the paths of the appropriate length that start in $S_4,S_5,S_6,T_4,T_5$, or $T_6$ and end in $u$. 
Furthermore, for every path that starts in $\{S_1,S_2,S_3,T_2,T_3\}$ and ends in $u$, there is a corresponding path with the same label that starts in $\{S_4,S_5,S_6,T_4,T_5,T_6\}$ and ends in $u$--this path can be obtained by replacing $S_1$ with $S_4$, $S_2$ with $S_5$, $S_3$ with $S_6$, $T_2$ with $T_5$ and $T_3$ with $T_6$. Finally, there are no paths of length at most $3$ from $Start$ to $u$. Hence, the terms in the column corresponding to $u$ in the matrix $\mathcal A_1^i$, $i\in \{1,2,3\}$, contain the labels for all paths of length $i$ that end in $u$. On the other hand, the terms in the diagonal element in this column  correspond to labels of the paths that start and end in $u$. 

It thus follows that to check \eqref{prop}, we need to verify that the nonzero terms in the non-diagonal elements of each column also appear in its diagonal element. For $\mathcal A_1$ and $\mathcal A_1^2$, this can be easily done by observing the matrices. For example, the entry $(3,3)$ of $\mathcal{A}_1^2$ equals $z^2+yz$ and contains all terms appearing in column 3 of $\mathcal{A}_1^2$, which are $yz$ and $z^2+yz$. We verified using a computer that $\mathcal A_1^3$ also satisfies the same condition. Hence, we have shown that all paths of length at most 3 ending in $\{S_4,S_5, S_6, T_4, T_5, T_6\}$ are duplicable.


{\it Part 2} : Now, we prove that all paths of length at most 3 ending in $\{S_1, S_2, S_3, T_2, T_3\}$ are duplicable. We first show that (\ref{prop}) holds for all states $\in \{S_1, S_2, T_2, T_3\}$ for paths of length  $\le3$, and also holds for $S_3$ for paths of length $1$ and $2$. Next, we show that while (\ref{prop}) does not hold for paths of length $3$ for $S_3$, all length-3 paths ending in $S_3$ are still duplicable. 

 Observe that there is no path of any length from any state $\in \{S_4, S_5, S_6, T_4, T_5, T_6\}$ to any state $\in \{Start, S_1, S_2, S_3,T_2,T_3\}$, hence we only need the (labeled) adjacency matrix $\mathcal A_2$ of the subgraph induced by $\{Start,S_1, S_2, S_3, T_2, T_3\}$. We have 
\begin{equation*}
\mathcal{A}_2 = 
\left[\begin{smallmatrix}
0 & x & 0 &0 &0&0\\
0&x & y & 0 & 0 & 0\\
0&0 & y & z & x & 0\\
0&0 & 0 & z & 0 & y\\
0&0 & y & 0 & x & 0\\
0&0 & 0 & z & 0 & y\\
\end{smallmatrix}\right],
\end{equation*}
\begin{equation*}
\mathcal{A}_2^2 = 
\left[\begin{smallmatrix}
0&x^2&xy&0&0&0\\
0&x^2 & xy & yz & yx & 0\\
0&0 & y^2 + xy & z^2 + yz & x^2 + yx & zy\\
0&0 & 0 & z^2+yz & 0 & y^2 + zy\\
0&0 & y^2+xy & yz & x^2 + yx & 0\\
0&0 & 0 & z^2+yz & 0 & y^2 + zy \\
\end{smallmatrix}\right],
\end{equation*}
where rows and columns correspond to $Start$, $S_1$, $S_2$, $S_3$, $T_2$, $T_3$, in that order. We observe that in $\mathcal{A}_2$ and $\mathcal{A}_2^2$, in each of the columns corresponding to $S_1$, $S_2$, $S_3$, $T_2$, and $T_3$, the terms in the diagonal entry contain the terms appearing in that column, implying that \eqref{prop} holds for all $u\in \{S_1, S_2, S_3, T_2, T_3\}$ and $j\in\{1,2\}$, i.e.\ for paths of length $1$ and $2$. By computing $\mathcal{A}_2^3$ using a computer, it can be checked that (\ref{prop}) holds for all states $u\in \{S_1, S_2, T_2, T_3\}$ for paths of length 3 as well.  

For $S_3$, there is a length-3 path $S_1-S_1-S_2-S_3$ with label $012$, for which there does not exist a corresponding path with the same label which starts and ends in $S_3$. Due to this fact (\ref{prop}) does not hold for $S_3$ for paths of length $3$. But for this length-$3$ path, we can traverse $S_3-S_4-S_5-S_6$ which also has label $012$. Now, since $S_6$ is a superstate of $S_3$, the path $012$ starting in $S_1$ and ending in $S_3$ is duplicable. The other length-3 paths ending in $S_3$ are $112$, $122$, $222$ and $212$. For each of these 4 paths, there exists a corresponding path with the same label that starts and ends in $S_3$ (see Figure \ref{fig1}). Hence, all length-3 path ending in $S_3$ are duplicable. This completes the proof of $S\subseteq L_R$.
 

Now that we have shown $S = L_R$, we use the Perron-Frobenius Theory~\cite{Immink,Lind} to count the number of sequences which can be generated via this deterministic finite automaton. We calculate the maximum absolute eigenvalue $e^*$ of the (unlabeled) adjacency matrix $B$ of the strongly connected component
of the finite automaton in Figure \ref{fig1} (i.e.\ the subgraph induced by $S_4, S_5, S_6, T_4, T_5, T_6$). The matrix $B$ can be obtained by replacing $x$, $y$, and $z$ in $\mathcal A_1$ by $1$,
\begin{equation*}
B =
\left[\begin{smallmatrix} 
1 & 1 & 0 & 1 & 0 & 0 \\
0 & 1 & 1 & 0 & 1 & 0 \\
1 & 0 & 1 & 0 & 0 & 1 \\
1 & 0 & 0 &1 & 0 & 0 \\
0 & 1 & 0 & 0 & 1 & 0 \\
0 & 0 & 1 & 0 & 0 & 1 \\
\end{smallmatrix}\right].
\end{equation*}
The maximum absolute eigenvalue of $B$ is $e^* = \frac{3+\sqrt5}{2}\simeq 2.618034$. By the Perron-Frobenius Theory,  $\capc(S) = \log_3 e^* \simeq  0.876036$.
\end{IEEEproof}
While the proof of the preceding theorem providing the exact capacity of the system under study is somewhat involved, it is easy to see why the capacity is strictly less than 1. One can observe from the regular expression for the finite automaton that it cannot generate a string which has $210$, $021$ or $102$ as a substring, implying that the system is not fully expressive. As we will see in Lemma~\ref{lem:expcap}, such systems cannot have capacity~$1$. It is worth noting that the set of strings that avoid $210$, $021$, and $102$ can be shown to have capacity $\simeq 0.914838$, which is slightly larger than the capacity of the system of the theorem.

\subsection{Expressiveness}
We now turn to study the expressiveness of tandem duplication systems with bounded duplication length. For completeness we start with binary systems, which is indeed the simplest case.
\begin{lem}\label{lem:bin}
	The system $S = \left(\{0,1\}, s, \mathcal{T}_{\leq 1}^{tan}\right)$, for any $s$ is not fully expressive.
\end{lem}
\begin{IEEEproof}
	The system cannot generate  ${(01)}^m$ as a substring of any string in $S$ for $2m>|s|$.
\end{IEEEproof}
As shown in Example~\ref{exm:1}, to obtain fully expressive binary systems, it suffices to increase the maximum duplication length to 2.

The next theorem is concerned with the expressiveness of $S = (\{0,1,2\}, s, \mathcal{T}^{tan}_{\leq 3})$. Larger alphabets and larger duplication lengths are considered in Theorems~\ref{thm:3}~and~\ref{thm:4}.
\begin{thm}	\label{thm:2}
	Consider $S = (\{0,1,2\}, s, \mathcal{T}^{tan}_{\leq 3})$, where $s$ is any arbitrary starting string $s$ $\in \{0,1,2\}^*$. Then, $S$ is not fully expressive. 
\end{thm} 
\begin{IEEEproof}
A $k$-\textit{irreducible} string is a string that does not have a tandem repeat $\alpha\alpha$, such that $|\alpha| \leq k.$ For example, $01201$, $01210$, $02101$, and $01210121$ are $3$-irreducible strings, while $01212$, $021021$ and $01112$ are not $3$-irreducible. To prove the theorem, we identify certain properties in new $3$-irreducible strings that may appear after a duplication and then construct a 3-irreducible string that is neither a substring of $s$, nor it satisfies the properties that every new 3-irreducible substring must satisfy. 

Consider a duplication event that transforms a sequence $z=uvw$ to $z^*=uvvw$, where $|v| \leq 3$. Let $x$ be a 3-irreducible string of length at least 4 that is present in $z^*$ but not in $z$. The string $x$ must intersect with both copies of $v$ in $z^*$ or else it is also present in $z$. Furthermore, it cannot contains $vv$, since otherwise it would not be 3-irreducible.  To determine the properties of $x$, we consider three case: $|v|=1,2,3.$ In what follows assume $a_1,a_2,a_3\in\Sigma$.

First, suppose $|v|=1$, say $v=a_1$. In this case, a string $x$ with the aforementioned properties does not exist as all new substrings contain the square $a_1a_1$.

Second, assume $|v|=2$, say $v=a_1a_2$. Then $z^*=ua_1a_2a_1a_2w$ and $x$ either ends with $a_1a_2a_1$ or starts with $a_2a_1a_2$.

Third, suppose $|v|=3$, say $v=a_1a_2a_3$. So $z^*=ua_1a_2a_3a_1a_2a_3w$. Recall that $|x|\ge4$. The string $x$ either ends with $a_1a_2a_3a_1$ or $a_2a_3a_1a_2$, or starts with $a_2a_3a_1a_2$ or $a_3a_1a_2a_3$.

So for any new 3-irreducible substring $x=x_1\dotsm x_j$, $x_i\in\Sigma$, $j\ge4$, we have $x_1=x_3$, $x_1=x_4$, $x_j=x_{j-2}$, or $x_j=x_{j-3}$. Now consider the string $(0121)^\ell0$, where $\ell>|s|$. This sequences is 3-irreducible but does not satisfy any of the 4 properties stated for $x$. Since it is not a substring of $s$ and it cannot be generated as a new substring, it is not a substring of any $y\in S$.
\end{IEEEproof}

Next we consider the system $\left(\Sigma, s, \mathcal{T}^{tan}_{\leq k}\right)$, $|\Sigma|\ge4$ in Theorem~\ref{thm:3}. The proof of the theorem, uses the following lemma, which states that the expressiveness of a system also has a bearing on its capacity.

\begin{lem}\label{lem:expcap}
	If a string system $S$ with alphabet $\Sigma$ is not fully expressive, then $\capc(S)<1$.
\end{lem}
\begin{IEEEproof}
	Since $S$ is not fully expressive, there exists a $z \in \Sigma^*$ that does not appear as a substring of any $y \in S$. Let $|z| = m$ and $\mu = n -m\lfloor{\frac{n}{m}}\rfloor$. We have $$|S\cap\Sigma^n| \leq {(|\Sigma|^m - 1)}^{\left \lfloor{\frac{n}{m}}\right \rfloor}{|\Sigma|}^\mu.$$ Since $m$ is finite, $\capc(S) < 1$.
\end{IEEEproof}
\begin{thm}
	\label{thm:3}Consider $S = \left(\Sigma, s, \mathcal{T}^{tan}_{\leq k}\right)$, where $|\Sigma| \geq 4$, $s$ is any arbitrary seed $\in \Sigma^*$ and $k$ is some finite natural number, then $S$ is not fully expressive, which also implies $\capc (S) < 1$.  
\end{thm}
\begin{IEEEproof}
Suppose $z=uvw\in S$, where $|v|\le k$, and let $z^*=uvvw$ be the result of a duplication applied to $z$. Furthermore, suppose that $x=x_1\dotsm x_j$, where $x_i\in\Sigma$ and $j>k$, is a square-free substring of $z^*$ but not $z$. Similar to the proof of Theorem~\ref{thm:2}, $x$ intersects both copies of $v$ but does not contain both. As a result, either $x_1=x_{1+i}$ or $x_j=x_{j-i}$,  for some $2\le i\le k$. 

For definiteness assume $\Sigma$ contains the symbols $\{0,1,2,3\}$. The sequence $0t0$, where $t$ is a square-free sequence over the alphabet $\{1,2,3\}$ and $|t|>\max\{|s|,k\}$, is not a substring of $s$ and cannot be generated as a substring since it does not satisfy the conditions stated for $x$ above. Note that such a $t$ exists since as shown by Thue~\cite{Thue}, for an alphabet size $\geq 3$, there exists a square-free string of any length. 
Hence $S$ is not fully expressive. The second part of the theorem follows from Lemma~\ref{lem:expcap}.
%
%
\end{IEEEproof}

 \begin{thm}
 	\label{thm:4}Consider $S = (\{0,1,2\}, 012,\mathcal{T}^{tan}_{\leq 4})$, then $S$ is a fully expressive string system. 
 \end{thm}
\begin{IEEEproof}
Let $S'=\left(\{0,1,2\},012,\mathcal T_{\le3}^{tan}\right)$. Clearly, $S'\subseteq S$. From the proof of Theorem~\ref{thm:1}, we know that the automaton of Figure~\ref{fig1} gives the same language as $S'$. By checking this automaton, we find that all strings of lengths 1, 2, and 3, except $021$, $210$, and $102$, appear as a substring of some string in $S'$ and, as a result, some string in $S$.
%
%
To generate $021$, $210$, and $102$ as substrings of some string in $S$, we proceed as follows: 
$$ 012 \rightarrow 0\underline{1212} \rightarrow \underline{01
{\bf 210}121}2 $$
\vspace{-1.5em}
$$ 012 \rightarrow \underline{012012} \rightarrow 01\underline{2020}12 \rightarrow
 0\underline{12{\bf 021}202}012$$
 \vspace{-1.5em}
$$ 012 \rightarrow \underline{012012} \rightarrow 01\underline{2020}12 \rightarrow 012\underline{020{\bf 102}01}2 $$ 
where the repeats are underlined.

We have shown that all strings of length 3 appear in $S$ as substrings. Now we show the same for every string $w=w_1w_2w_3w_4$ of length 4. To do so, we study 3 cases based on the structure of $w$:

I) First, suppose that $w_4$ is the same as $w_1$, $w_2$, or $w_3$. For generating such $w$ as a substring,  we first generate $w'=w_1w_2w_3$ as a substring of some string and then do a tandem duplication of $w_3$ if $w_4 = w_3$, of $w_2w_3$ if $w_4 = w_2$ and of $w_1w_2w_3$ if $w_4 = w_1$.

II) Suppose I) does not hold but $w_1=w_2$ or $w_2=w_3$. If the former holds, first generate $w_1w_3w_4$ and then duplicate $w_1$, and if the latter hold, generate $w_1w_2w_4$ and duplicate $w_2$.

III) If neither I) nor II) holds, then $w=1210$, up to a relabling of the symbols. In this case, we first generate $w'=0121$ and then do a tandem duplication of $w'$ to get $w$. Note that $w'$ is of type considered in I).

Until now, we have shown that all strings $w$ of length at most $4$ appear as a substring of some string in $S$. We use induction to complete the proof. Suppose all strings of length at most $m$ appear as a substring of some string in $S$, where $m\geq 4$. We show that the same holds for strings of length $m+1$.

Consider an arbitrary $w = a_1a_2\dotsm a_ma_{m+1}$. We now consider two cases:

i) If all three letters in the alphabet occur at least once in $a_{m-3}a_{m-2}a_{m-1}a_m$, then $a_{m+1}$ equals $a_{m-3}$, $a_{m-2}$, $a_{m-1}$, or $a_m$, and $w$ can be generated as a substring by a tandem duplication of some suffix of size $\leq 4$ of $w' = a_1a_2\dotsm a_m$. Note that by the induction hypothesis $w'$ can be generated as a substring of some string.

ii) If at least one letter in the alphabet does not occur in $a_{m-3}a_{m-2}a_{m-1}a_m$, then $a_{m-3}a_{m-2}a_{m-1}a_m$ is a sequence over binary alphabet and so it has a tandem repeat. Therefore $w$ can be generated as a substring by tandem duplication. Hence, we have proved the Theorem. 
\end{IEEEproof}

Table \ref{table:div2} summarizes the result of this subsection. It can be observed from the table that a change of behavior in expressiveness occurs when the size of the alphabet increases to 4. If the size of the alphabet is $1$, $2$, or $3$, for sufficiently large maximum duplication length, the systems are fully expressive. However, if the size of the alphabet is at least $4$, then regardless of the maximum duplication length, the system is not fully expressive. This change is related to the fact that for alphabets of size $1$ and $2$, all square-free strings are of finite length, but for alphabets of size $3$ and larger, there are square-free strings of any length. Specifically, in case ii) in the proof of Theorem~\ref{thm:4}, we used the fact that the binary string $a_{m-3}a_{m-2}a_{m-1}a_m$ has a tandem repeat. To adapt this proof for $|\Sigma|\ge4$, we would need to show that the $|\Sigma|-1$-ary string $a_{m-3}a_{m-2}a_{m-1}a_m$ has a tandem repeat. But this is not in general true, since there are square-free strings over alphabets of size at least 3 per Thue's result~\cite{Thue} and indeed we showed in Theorem~\ref{thm:3}, again using Thue's result, that the system $\left(\Sigma,s,\mathcal T_{\le k}^{tan}\right)$ is not fully expressive for $|\Sigma|\ge4$ and any $k$.

\begin{table}
	\begin{center}
		\begin{tabular}{|c|c|c|c|c|}
			\hline
			$\Sigma$ & $s$ &$k$ & fully expressive & Reason \\\hline\hline
			$\{0\}$ & $0$ & $\geq 1$ & Yes & Trivial \\\hline
			$\{0,1\}$ & arbitrary & $ 1$ & No & Lemma~\ref{lem:bin} \\\hline 
			$\{0,1\}$ & $01$ & $\geq 2$&  Yes & Example~\ref{exm:1}\\\hline
			$\{0,1,2\}$ & arbitrary & $\leq 3$ &No & Theorem~\ref{thm:2}\\\hline
			$\{0,1,2\}$&$012$& $\geq 4$& Yes& Theorem~\ref{thm:4}\\\hline
			$|\Sigma|\geq 4$ &arbitrary &arbitrary & No & Theorem~\ref{thm:3} \\\hline
		\end{tabular}
	\end{center}
	\caption{Expressiveness of tandem duplication string systems $(\Sigma, s, \mathcal{T}^{tan}_{\leq k}$).}
	\vspace{-1.5em}
	\label{table:div2}
\end{table}
\section{Regular Languages for Tandem Duplication String Systems}\label{sec:reg}
Regular languages for tandem duplication string systems are easier to study due to the fact that one can use tools from Perron-Frobenius theory~\cite{Immink, Lind} to calculate capacity. It was proved in \cite{Leupold2} that for $|\Sigma| \geq 3$ and maximum duplication length $\geq 4$, the language defined by tandem duplication string systems is not regular, if the seed contains $abc$ as a substring such that $a,b$ and $c$ are distinct. However, if the maximum duplication length is 3, this question was left unanswered. In Theorem~\ref{thm:5}, we show that the language resulting from a tandem duplication system with the maximum duplication length of 3 is regular regardless of the alphabet size and seed. Further, in Corollary~\ref{cor:1} we 	 characterize the exact capacity of such tandem duplication string systems. 

\begin{thm}
	\label{thm:5}Let $S = (\Sigma, s, \mathcal{T}^{tan}_{\leq 3})$, where $\Sigma$ and $s$ are arbitrary. The language defined by $S$ is regular.
\end{thm}
\begin{IEEEproof} 
 We first assume that $s=a_1\dotsm a_m$, where $a_i$ are distinct. The case in which $a_i$ are not distinct is handled later.

\newcommand{\blockTwo}[2]{{\left(#1^+#2^+\right)}}
\newcommand{\blockThree}[3]{{B_{#1#2#3}}}
For $3\le j\le m$, let 
\begin{equation*}
\begin{split}
R_{a_1\cdots a_j} = a_1^+a_2^+ {\blockTwo{a_1}{a_2}}^*
&a_3^+\blockTwo{a_2}{a_3}^*\blockThree{a_1}{a_2}{a_3}^*\\
&a_4^+\blockTwo{a_3}{a_4}^*\blockThree{a_2}{a_3}{a_4}^*\\
&\cdots \\
&a_i^+\blockTwo{a_{i-1}}{a_i}^*\blockThree{a_{i-2}}{a_{i-1}}{a_i}^*\\
&\cdots\\
&a_j^+\blockTwo{a_{j-1}}{a_j}^*\blockThree{a_{j-2}}{a_{j-1}}{a_j}^*,
\end{split}
\end{equation*} 
where, for $a,b,c\in\Sigma$, 
\begin{equation*}
\begin{split}
\blockThree abc &=  {a^+{(c^+a^+)}^*b^+{(a^+b^+)}^*c^+{(b^+c^+)}^*}.
\end{split}
\end{equation*}


We already know from Theorem~\ref{thm:1} that $S = (\Sigma, s, \mathcal{T}^{tan}_{\leq 3})$ with $s=a_1\dotsm a_m$ is a regular language if $m=3$. We show that for $m \geq 4$, $S$ represents a regular language whose regular expression is given by  $R_{a_1a_2\cdots a_m}$. Let $L_R$ be the language defined by $R_{a_1a_2\cdots a_m}$.  It suffices to show $L_R=S$.

 \begin{table*}[t]
 	\label{cap_tab_1}
 	\begin{center}
 		\begin{tabular}{|c|c|c|c|}
 			\hline
 			$\Sigma$ & $s$ &$k$ & Capacity\\\hline 
 			$\{0,1\}$ & $01$ & $1$& $ 0$\\\hline
 			$\{0,1\}$ & $01$ & $\geq 2$ &$ 1$\\\hline
 			arbitrary& arbitrary but not $a^m$ for some $a~\in~\Sigma$ & $2$ &$\log_{|\Sigma|} 2$\\\hline
 			$\{0,1,2\}$ &$012$ &$3$ & $ \log_3 \frac{3+\sqrt{5}}{2}$ \\\hline
 			arbitrary   & $xabcy$ ($x$ and $y~ \in ~\Sigma^*$, $a, b$ and $c ~\in ~\Sigma$ and $a\neq b \neq c\neq a$)&$3$&$\log_{|\Sigma|}\frac{3+\sqrt{5}}{2}$\\\hline
 			arbitrary & No $3$ consecutive symbols in the seed are all distinct and $s \neq a^m$ for $a \in \Sigma$ & $3$ & $\log_{|\Sigma|} 2$\\\hline
 		\end{tabular}
 	\end{center}
 	\caption{Capacity values for different tandem duplication string systems $(\Sigma, s,\mathcal{T}^{tan}_{\leq k})$.}
 	\vspace{-1.5em}
 	\label{table:cap2}
 \end{table*}

We first show that $L_R \subseteq S$ by proving $R_{a_1a_2\cdots a_m} \dd{\le3}s$. To do so, we show by induction that $R_{a_1a_2\dotsm a_i}\dd{\le3}a_1a_2\dotsm a_i$. First note that this holds for $i=3$, from the proof of Theorem~\ref{thm:1}. Assuming that it holds for $i$, to show that this also holds for $i+1$, where $i\ge3$. We write
\begin{equation*}
\begin{split}
R_{a_1a_2\dotsm a_{i+1}}
&\dd{\le3}R_{a_1a_2\dotsm a_{i}}a_{i+1}^+\blockTwo{a_i}{a_{i+1}}^*\blockThree{a_{i-1}}{a_i}{a_{i+1}}^*\\
&\dd{\le3}a_1a_2\dotsm a_i a_{i+1}\left(a_{i}a_{i+1}\right)^*\blockThree{a_{i-1}}{a_i}{a_{i+1}}^*\\
&\dd{\le3}a_1a_2\dotsm a_i a_{i+1}\left(a_{i-1}a_{i}a_{i+1}\right)^*or \\& \quad\quad\quad a_1a_2\dotsm a_i a_{i+1}\left(a_{i-1}a_{i+1}a_{i-1}a_{i}a_{i+1}\right)^*\\
&\dd{\le3}a_1a_2\dotsm a_i a_{i+1}.
\end{split}
\end{equation*} 
Here we have used the fact that $c\blockThree{a}{b}{c}\dd{\le3}cabc$ which follows from  \eqref{eq:B2}. Hence, $L_R \subseteq S$.

We now show that $ S \subseteq L_R$. Note that the seed $s$ is in $L_R$. It thus suffices to show that if $x = pqr \in L_R$, then $y = pq^2r \in L_R$, where $p, q,r\in\Sigma^*$ and $|q| \leq 3$.  We have the following five cases:
\begin{enumerate}
\item $q = b$, $q=bb$ or $q=bbb$, for some $b\in \Sigma$: Since each symbol in the regular expression $R_{a_1\dotsm a_m}$ is followed by a $+$ or $*$ as a superscript, if $q$ represents a run and $pqr\in L_R$, then so is $pq^2r$. 
\item $q = bc$ for distinct $b,c\in \Sigma:$  Here $q$ represents a length-$2$ path in the finite automaton for a regular expression of the form ${\blockTwo{b}{c}}^*$, $b^+{\blockTwo{c}{b}}^*$, $b^+\blockThree{c}{a}{b}$, $\blockThree{a}{b}{c}$, $\blockThree{b}{c}{a}$, $\blockThree{b}{a}{c}$, $\blockThree{c}{a}{b}$, $\blockThree{a}{c}{b}$ or $b^+c^+$. We know from the proof of Theorem \ref{thm:1} that $bc$ is duplicable in ${\blockTwo{b}{c}}^*$, $\blockThree{a}{b}{c}$, $\blockThree{b}{c}{a}$, $\blockThree{b}{a}{c}$, $\blockThree{c}{a}{b}$ and $\blockThree{a}{c}{b}$. For $b^+{\blockTwo{c}{b}}^*$ and $b^+\blockThree{c}{a}{b}$, we enter a state in the finite automaton for ${\blockTwo{c}{b}}^*$ and $\blockThree{c}{a}{b}$ respectively with incoming edge labeled by $c$. In this state, we can again duplicate path $bc$ and return back to the same state.

 The finite automaton for $b^+c^+$ is followed by the finite automaton for ${\blockTwo{b}{c}}^*$, so $bc$ can be duplicated in the automaton for ${\blockTwo{b}{c}}^*$. The duplicate $q = bc$ generated here in ${\blockTwo{b}{c}}^*$ ends in some state $C$ which is a superstate of the state $D$ in which the original $q$ in $pqr$ ended. Since $C$ is a superstate of $D$, $r$ can also be generated from $C$. Hence $pq^2r \in L_R$.
\item $q = bbc$ or $bcc$ for distinct $b, c \in \Sigma:$ Here $q$ represents a length-3 path. We only consider $q=bbc$; the other case is similar. If $pbbcr\in L_R$, then $pbcr\in L_R$ as well, since every symbol in $R_{a_1\dots a_m}$ is followed by a $+$ or $*$ as a superscript. Now we already know from case 2 above if $pbcr$ can be generated then $pbcbcr$ can also be generated. Now from case 1 above, we also know if $pbcbcr$ can be generated then $pbbcbcr$ can also be generated. Further using case 1 again, we can generate $pbbcbbcr$ from $pbbcbcr$. Hence $pq^2r \in L_R.$
\item $q = abc$ for distinct $a, b, c \in \Sigma:$ Here $q$ represents a length-$3$ path in the finite automaton for $\blockThree{\sigma(a}{b}{c)}$ ($\sigma(abc)$ represents any permutation of $a,b,c$), $a^+{\blockTwo{b}{c}}^*$, $a^+\blockThree{b}{c}{a}$, ${\blockTwo{a}{b}}^*c^+$, $\blockThree{d}{a}{b}c^+$, ${\blockTwo{a}{b}}^*\blockThree{c}{a}{b}$ or $a^+b^+c^+$. We know from the proof of Theorem \ref{thm:1} that $abc$ is duplicable in $\blockThree{\sigma(a}{b}{c)}$. The same reasoning holds for $a^+\blockThree{b}{c}{a}$ and ${\blockTwo{a}{b}}^*\blockThree{c}{a}{b}$.

 The finite automaton for $a^+{\blockTwo{b}{c}}^*$, ${\blockTwo{a}{b}}^*c^+$, $\blockThree{d}{a}{b}c^+$ and $a^+b^+c^+$ is followed by a finite automaton for $\blockThree{a}{b}{c}$, so $q$ can be duplicated in the finite automaton for $\blockThree{a}{b}{c}$. The duplicate $q$ ends in some state $E$ which is the superstate of the state $F$ in which the original $q$ in $pqr$ ended. Since, $E$ is a superstate of $F$, therefore $r$ can also be generated from $E$. Hence $pq^2r  \in L_R$. 

\item $q = cbc$ for distinct $b,c \in \Sigma:$ Here $q$ represents a length-$3$ path that can be generated by the finite automaton for  ${\blockTwo{c}{b}}^*$, ${\blockTwo{b}{c}}^*$, $\blockThree{\sigma(c}{b}{a)}$, $c^+{\blockTwo{c}{b}}^*$ or $c^+b^+{\blockTwo{c}{b}}^*$. We know from the proof of Theorem 1 that $cbc$ is duplicable in ${\blockTwo{c}{b}}^*, {\blockTwo{b}{c}}^*$ and $\blockThree{\sigma(c}{b}{a)}$. As the state where $q$ in $pqr$ ends lies in the finite automata for either ${\blockTwo{c}{b}}^*, {\blockTwo{b}{c}}^* or \blockThree{\sigma(c}{b}{a)}$, it can be duplicated again the same finite automaton. The duplicate $q$ ends in the superstate of the state in which the original $q$ in $pqr$ ended. Hence $pq^2r \in L_R$.
\end{enumerate}
This completes the proof of $S\subseteq L_R$.

We have proved the statement of Theorem \ref{thm:5} assuming all $a_i$'s in the seed $s$ to be distinct. Now assume the symbols of $s$ are not distinct. We color the symbols of $s$ so that they become distinct and obtain the system $\tilde S=\left(\tilde\Sigma,\tilde s,\mathcal T_{\le3}^{tan}\right)$. Applying the preceding proof for distinct symbols to $\tilde S$, we find that $\tilde S$ is regular. Let $h:\tilde\Sigma\to\Sigma$ be a mapping that removes the colors. By \cite{Shallit}, we have that $S=h(\tilde S)$ is also regular. 
\end{IEEEproof}
An immediate corollary on the capacity of tandem duplication string system considered in Theorem \ref{thm:5} is
\begin{cor}\label{cor:1}
If for $S$ in Theorem~\ref{thm:5}, $s$ contains $abc$ as a substring such that $a, b,$ and $c \in\Sigma$ are distinct, then $\capc(S) = \log_{|\Sigma|} \frac{3+\sqrt{5}}{2} \simeq 0.876036\log_{|\Sigma|}3$. Otherwise, except for the seed of the form $a^m$, $\capc(S) = \log_{|\Sigma|}2$. If $s = a^m$, $\capc(S) = 0$.
\end{cor}
\begin{IEEEproof}
By the Perron-Frobenius Theory \cite{Immink}, \cite{Lind}, for a regular language $L_R$, the capacity is given by the log of the maximum eigenvalue of the adjacency matrix of the strongly connected components. In the case when $abc$ occurs as a substring of the seed $s$ such that $a, b$ and $c \in \Sigma$ are distinct, then the adjacency matrix of the finite automaton for $B_{abc}$ (strongly connected component of the finite automaton for $R_{a_1a_2\cdots a_m}$) has the maximum eigenvalue. Therefore, the $\capc(S) = \log_{|\Sigma|} \frac{3+\sqrt{5}}{2} \simeq 0.876036\log_{|\Sigma|}3$ (see proof of Theorem~\ref{thm:1} for the adjacency matrix). 

For the case when no $3$ consecutive symbols in the seed $s$ are all distinct and $s\neq a^m$, the maximum capacity component is a finite automaton only over 2 distinct symbols as in Figure \ref{fig000}. Hence the capacity is $\log_{|\Sigma|} 2$.

When seed $s = a^m$, there is at most one sequence of any given length in the system. Hence $\capc(S) = 0.$
\end{IEEEproof}

The following examples illustrate the statement of Theorem~\ref{thm:5} and an application of its proof method.
\begin{exm}
The string system $ S = (\{0,1,2,3\}, 0123, \mathcal{T}^{tan}_{\leq 3})$ is regular by Theorem~\ref{thm:5} and the regular expression is given by 
\begin{equation*}\label{reg_exp_4}
R_{0123} 
=  0^+1^+{(0^+1^+)}^*2^+{(1^+2^+)}^*{B_{012}}^*3^+{(2^+3^+)}^*{B_{123}}^*.
\end{equation*}
By Corollary~\ref{cor:1}, the capacity of this system $\simeq 0.876036\log_4 3 \simeq 0.694242.$
\myendexm\end{exm}
\begin{exm}
 The string system $S = (\{0,1,2\}, 0112, \mathcal{T}^{tan}_{\leq 3})$ is regular by Theorem~\ref{thm:5}, and the regular expression is given by
\begin{equation*}\label{reg_exp_4_alt}
R_{0112} 
= 0^+1^+{(0^+1^+)}^*1^+{(1^+1^+)}^*
{B_{011}}^*2^+{(1^+2^+)}^*{B_{112}}^*.
\end{equation*}
By Corollary~\ref{cor:1}, the capacity of this system is given by $\log_3 2 \simeq 0.63093.$
\myendexm\end{exm}

When $a_i$'s are assumed to be distinct it can be verified from the regular expression $R_{a_1\cdots a_j}$ in the proof of Thereom \ref{thm:5} that the last occurence of $a_i$ is before the first occurence of $a_{i+3}$ for any $i = 1,2,\dotsm, j-3$ for all $z \in S$. Motivated by this, we state the following lemma regarding the structure of words in tandem duplication systems with bounded duplication lengths

\begin{lem}\label{lem:1} Let $s = a_1\cdots a_m$, where $a_i\in\Sigma$ are distinct. Then for any $ z \in S = \left(\Sigma, s, \mathcal{T}^{tan}_{\leq k}\right)$ and any $i= 1,\dotsc , m-k$, the last occurrence of $a_i$ is before the first occurrence of $a_{i+k}$ and the gap between them is at least k-1 (not counting $a_i$ and $a_{i+k}$). 
\end{lem}
\begin{IEEEproof}
Fix the value of $i$. We prove the lemma by induction. Clearly, the lemma holds for $z=s$. Assuming that it holds for $x\in S$, we show that it also holds for $y=T(x)$ for any $T\in \mathcal T_{\le k}^{tan}$.

Assume $x=\alpha a_i \beta a_{i+k} \gamma$, where $\alpha,\beta,\gamma\in\Sigma^*$ and where $a_i$ and $a_{i+k}$ in this expression refer to the last occurrence of $a_i$ and the first occurrence of $a_{i+k}$ in $x$, respectively. Since, by assumption $|\beta|\ge k-1$, the tandem duplication $T$ cannot contain a substring that contains both the last occurrence of $a_i$ and the first occurrence of $a_{i+k}$. If the tandem duplication $T$ duplicates a substring of $\beta$, then the gap between the last $a_i$ and the first $a_{i+k}$ in $y$ is larger than that of $x$. In every other case, the gap stays the same. So the gap in $y$ is at least as large as the gap in $x$, which is $|\beta|\geq k-1$.
\end{IEEEproof}
The following example follows for maximum duplication length $2$ using the same idea as in Theorem \ref{thm:5}
\begin{exm}\label{exm:9}
The string system $ S = (\Sigma, a_1a_2\cdots a_m, \mathcal{T}^{tan}_{\leq 2})$ is regular. This can be proved using the same method as used in the proof of Theorem~\ref{thm:5}.  The regular expression $Q_{a_1a_2\cdots a_m}$ for $m \geq 2$ is given by
\begin{equation*}\label{reg_exp_T}
Q_{a_1a_2\cdots a_m} = a_1^+a_2^+{(a_1^+a_2^+)}^*a_3^+{(a_2^+a_3^+)}^*\cdots a_m^+{(a_{m-1}^+a_m^+)}^*.
\end{equation*}
\myendexm\end{exm}

\begin{figure}
\centerline{\includegraphics[width=3in,keepaspectratio]{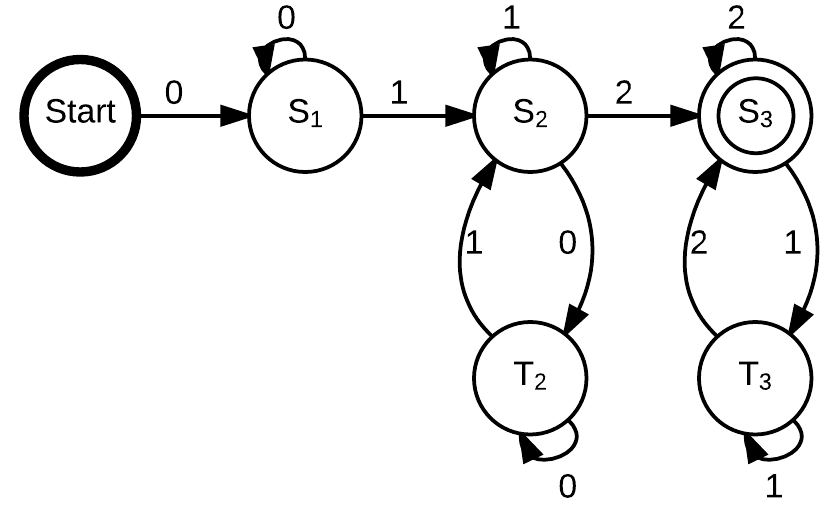}}
\caption{Finite automaton for $S = (\{0, 1, 2\}, 012,\mathcal{T}_{\leq 2}^{tan})$. The regular expression $R = 0^+1^+{(0^+1^+)}^*2^+{(1^+2^+)}^*.$}
\label{fig0}
\end{figure}

 The finite automaton for a special case of Example \ref{exm:9} with $|\Sigma| = 3$ is given in Figure \ref{fig0}. 
\begin{cor}\label{cor:2}
	The capacity for $S = (\Sigma, a_1a_2\cdots a_m, \mathcal{T}^{tan}_{\leq 2})$ is given by $\log_{|\Sigma|} 2$, except for the case in which seed $s = a^m$ for $a ~\in~\Sigma$. In that case, the capacity is $0$.
\end{cor}
\begin{IEEEproof}
As in Proof of Corollary~\ref{cor:1}, By the Perron-Frobenius Theory, for a regular language, the capacity is given by the log of the maximum eigenvalue of the adjacency matrix of the strongly connected components. Except for the case when seed $s = a^m$, for all other cases $ab$ ($a,b ~\in~\Sigma$) occurs as a substring of the seed $s$ such that $a\neq b$. Hence, the maximum capacity component in the finite automaton for $Q_{a_1a_2\cdots a_m}$ is ${(a^+b^+)}^+$ for which the capacity is $\log_{|\Sigma|} 2$.
%
\end{IEEEproof}

 Our capacity results are listed in Table \ref{table:cap2}.

\section{Conclusion}\label{sec:conc}
In this paper, we showed that for tandem duplication string systems with bounded duplication length if the maximum duplication length is $3$ or less, the language described by the string system is regular. Further, we computed exact capacities for these systems. As a future work, we would like to calculate capacities for bounded tandem duplication string systems with maximum duplication length greater than $3$. 

Using Thue's result~\cite{Thue}, we showed that a tandem duplication string system cannot be fully expressive if the alphabet size is $\geq4$. However, for an alphabet of size $3$ or less such systems can be fully expressive. This way, we completely characterized fully expressive and non-fully expressive tandem duplication string systems with bounded duplication length. As a future work, we would like to generalize the notion of expressiveness by counting the asymptotic number of \textit{substrings} of length $n$ that a string system can generate. Mathematically, we define the expressiveness $Exp(S)$ of a string system $S$ as
$$ Exp(S) = \limsup_{n\rightarrow \infty}\frac{\log_{|\Sigma|}E_n(S)}{n}.$$Here $E_n(S)$ represents the number of substrings of length $n$ that can be generated by $S$. It is notable here that with this definition of expressiveness, a fully expressive string system $S$ has $Exp(S) = 1$. 

In this paper, we looked at questions related to the generation of a diversity of sequences from a seed given a tandem duplication rule. One can also study the minimum number of steps required to deduplicate a given sequence of length $n$ to a squarefree seed and therefore define the notion of distance between a sequence and its seed given a tandem duplication rule. It is notable here that the same sequence can be deduplicated to more than one squarefree seed given a tandem duplication rule. For example: the sequence $012101212$ can be deduplicated to $012$ as well as $0121012$ under bounded tandem duplication with maximum duplication length $4 $ in the following way $$\underline{01210121}2 \dd{\le4} 0\underline{1212} \dd{\le4} 012.$$
$$01210\underline{1212} \dd{\le4} 0121012.$$
Here the underlined portion represents the repeat that is being deduplicated in a given step.




\end{document}